\documentclass[final,3p,times,twocolumn]{elsarticle}

\usepackage{amsmath}

\usepackage{chemformula}

\usepackage{xcolor}
\usepackage{comment}
\usepackage[range-units = single]{siunitx}
\usepackage[space]{grffile}

\usepackage{hyperref}
\hypersetup{colorlinks=true,linkcolor = blue}
\usepackage{cleveref}

\journal{Journal of Magnetic Resonance}

\begin{document}

\begin{frontmatter}



\cortext[cor1]{Corresponding author: \href{mailto:akk916@ic.ac.uk}{akk916@ic.ac.uk}}

\title{Pulsed Electron Spin Resonance of an Organic Microcrystal by Dispersive Readout}

\author[npl,icl]{Ailsa K. V. Keyser\corref{cor1}}
\author[npl]{Jonathan J. Burnett}
\author[ch]{Sergey E. Kubatkin}
\author[ch]{Andrey V. Danilov}
\author[icl]{Mark Oxborrow}
\author[npl]{Sebastian E. de Graaf}
\author[npl]{Tobias Lindstr\"om}
\address[npl]{National Physical Laboratory, Hampton Road, Teddington, TW11 0LW, UK}
\address[icl]{Imperial College London, Exhibition Road, SW7 2AZ, UK}
\address[ch]{Department of Microtechnology and Nanoscience, MC2, Chalmers University of Technology, SE-41296 Gothenburg, Sweden}

\begin{abstract}
We establish a testbed system for the development of high-sensitivity Electron Spin Resonance (ESR) techniques for small samples at cryogenic temperatures. Our system consists of a \ch{NbN} thin-film planar superconducting microresonator designed to have a concentrated mode volume to couple to a small amount of paramagnetic material, and to be resilient to magnetic fields of up to $\SI{400}{\milli\tesla}$. At $\SI{65}{\milli\kelvin}$ we measure high-cooperativity coupling ($C \approx 19$) to an organic radical microcrystal containing $10^{12}$ spins in a pico-litre volume. We detect the spin-lattice decoherence rate via the dispersive frequency shift of the resonator. Techniques such as these \textcolor{black}{could be suitable for applications in quantum information as well as for pulsed ESR interrogation of very few spins to provide insights into the surface chemistry of, for example, the material defects in superconducting quantum processors}. 
\end{abstract}



\end{frontmatter}


\section{Introduction}
\label{intro}
Electron Spin Resonance (ESR) is an essential technique for characterising a wide range of paramagnetic materials, 
 but its application to the study of micro-scale samples, as required for many emerging technologies, is beyond the reach of most commercially available systems \cite{Schweiger2001}. 
A conventional commercial ESR spectrometer \cite{Abe2011a} typically has a three-dimensional cavity with a low $Q\sim 10^3$ factor
and a large mode volume of $\sim \SI{100}{\milli\metre^3}$ through which microwave energy is weakly coupled to the 
(thereby necessarily large number of) electron spins \cite{Morton2018}.
ESR spectrometers which use as the cavity a superconducting planar microresonator benefit from their combination of higher $Q$ factors $>\num{e5}$ \emph{and} smaller mode volumes $\sim \SI{e-4}{\milli\metre^3}$, as are required to detect small volumes of paramagnetic material \cite{Blank2013,BienfaitThesis}. 

Developments in high-sensitivity ESR 
have occurred in tandem with those made by quantum engineers intent on measuring weak microwave signals from superconducting qubits \cite{Krantz2019}. Resulting milestones in this new regime of ESR equipment include 
quantum-limited Josephson Parametric Amplifiers which add no more than the minimum number of noise photons to the signal from the electron spins \cite{Yaakobi2013,Zhou2014}; Arbitrary Waveform Generators (AWGs) able to shape pulses at picosecond resolution; and high $Q$ superconducting planar microresonators resilient to $>\nolinebreak\SI{1}{\tesla}$ magnetic field and capable of producing a uniquely small mode volume for more sensitive ESR experiments at cryogenic temperatures \cite{Morton2018,Mahashabde2020,Samkharadze2016}. So far, a combination of such devices has pushed absolute ESR single-shot spin sensitivity to the inductive-detection of fewer than \num{e3} spins \cite{Eichler2017,Probst2017}. 

As such, it is now becoming possible to measure ultra-small volumes of paramagnetic samples for the first time \cite{Ranjan2020}. This will enable, for instance, the characterisation and chemical identification of individual nanoscale objects, or small spin clusters where anisotropy in the spin Hamiltonian otherwise makes characterisation of the material via large mode volume detectors challenging. Such proof-of-principle experiments also constitute an essential route towards unlocking the use of large numbers of self-assembled molecular spin clusters as high coherence elements for quantum computing \cite{GaitaArino2019,Ghirri2016}.

Most conventional ESR methodology pertains to the regime of a large number $N > \num{e13}$ of electron spins \textcolor{black}{from which the goal is spin detection and characterisation \cite{Schweiger2001}. Our resonator is for use in the contrasting limit of ESR for small volumes of paramagnetic material. Resonant detection of fewer than \num{100} spins has recently been achieved \cite{Eichler2017,Probst2017} using high Q superconducting resonators, however, detection is limited to spins with coherence times well exceeding the cavity lifetime. The opposite limit of ESR for materials characterisation of a small number of spins with poor coherence times has yet to be demonstrated, and new approaches are needed to address this challenge. Further to this, quantum computing applications using electron spins would require detection of the quantum state of the spins. For this purpose,} 
 care must be taken to avoid significant coherent energy exchange between spins and the interrogating cavity. This back-action means that readout of the cavity probe can induce a change in the projected spin states, whereas an ideal repeated measurement should not disturb the states of the measured observable, a concept known as a quantum non-demolition (QND) measurement.  To this end new protocols need to be developed for this regime of ESR \textcolor{black}{measurement of small numbers of spins, even with short coherence times, and for quantum computing applications, in order to fully exploit the potential of coherent information storage in spin ensembles. We lastly highlight that the readout of quantum states from high density spin ensembles allows the coupling to be enhanced by factor $\sqrt{N}$ whilst the small volume is beneficial in reducing field inhomogeneity and so implementing coherent control pulses.}

Here we demonstrate the use of dispersive readout, a technique commonly used for QND measurements of qubits \cite{Wallraff2004}, which allows the resonator to act as a probe of the spin polarisation when the two systems are detuned. In pulsed ESR it has been previously deployed for measurements of macroscopic paramagnetic samples, including Nitrogen-Vacancy (NV) \cite{Amsuss2011, Ebel2020} or substitutional Nitrogen centres in diamond \cite{Ranjan2013}, ferromagnetic resonance \cite{Haigh2015} and the dispersive detection of NV centres via a transmon qubit \cite{Hu2017}. 
We apply dispersive readout instead to 
a pico-litre volume of paramagnetic material: a single microcrystal of an organic radical \cite{Christensen2014} which we characterise here at \SI{}{\milli\kelvin} temperatures. The necessarily small mode volume is achieved with a specially-designed high $Q$ superconducting resonator, which can be magnetically coupled to any small solid-state sample (here the $\SI[product-units = single]{70x6x6}{\micro\metre^3}$ microcrystal) as long as it is carefully aligned within the resonator mode volume. 

We present continuous wave (CW) data characterising the high-cooperativity resonator-spin coupling, and use pulsed microwave measurements to measure the spin-lattice decoherence time $T_1$ dispersively. 
Whereas previous dispersive readout ESR measurements \cite{Amsuss2011,Ebel2020,Ranjan2013} used the resonator to both excite the spins and detect their response, we have two separate microwave lines: one to excite the spins and another to readout the resonator. This enables us to probe the resonator independently of the spin ensemble. While this additional freedom is unusual in ESR experiments, it allows us to explore the resonator dynamics as the spins decay to their ground state.

\section{Materials \& methods}
\label{mnm}
\subsection{Microresonator setup}

\begin{figure}[t]
	\includegraphics[width=\linewidth]{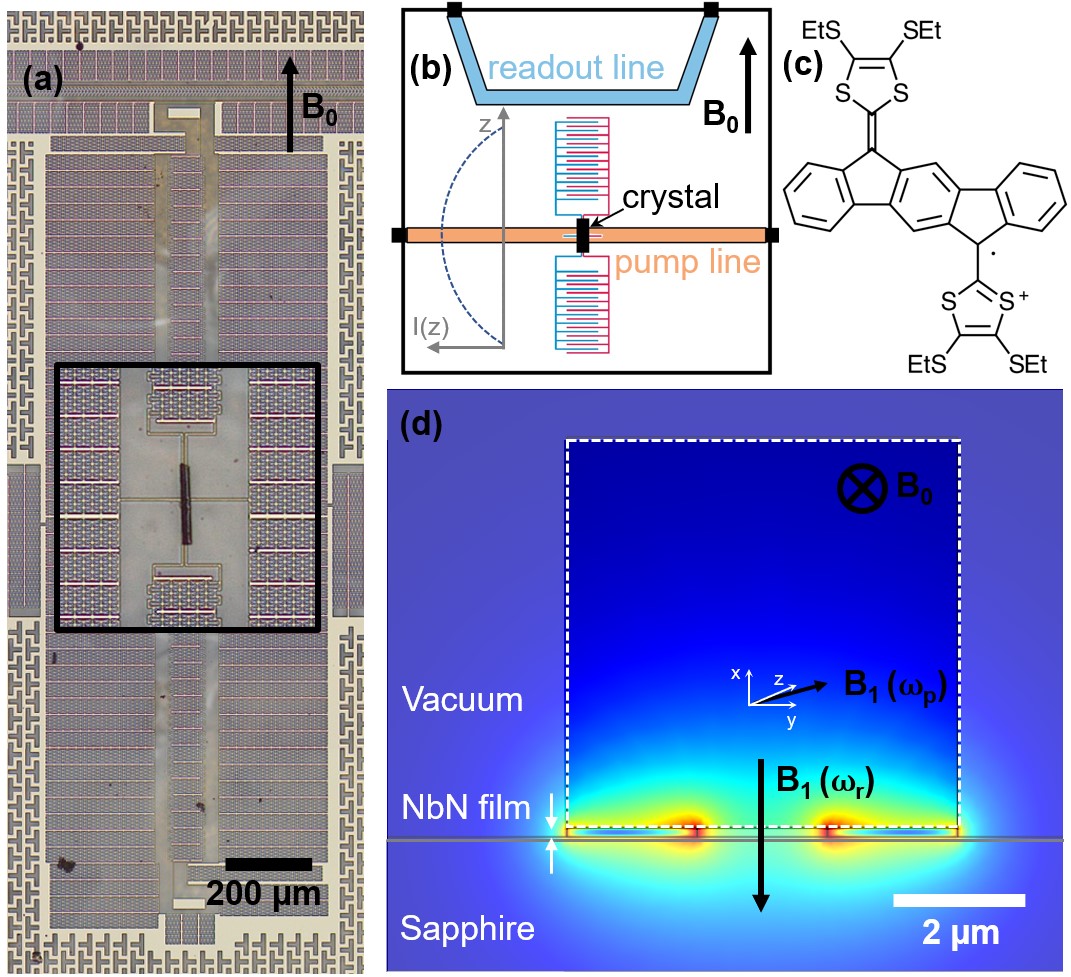}
	\caption{(a) Details of the fractal resonator design showing the crystal alignment (inset panel is double magnification compared with the rest of the image) and (b) the layout of the chip with a schematic of the resonator and a (co-planar) transmission line used to readout the resonator (coloured blue) and a pump line (coloured orange) on the PCB \SI{330}{\micro\metre} below the plane of the resonator. A single (black) microcrystal consisting of the molecular radical shown in (c) is aligned at the magnetic anti-node of the resonant mode as shown by the microwave current distribution $I(z)$ along the resonator length $z\parallel B_0$. The magnitude of the microwave magnetic field $B_1(\omega_r)$ generated by the resonator, in the plane normal to the $z$-axis, is shown by colour gradient in the \textsc{Comsol} simulation (d), and the approximate contour of the microcrystal has been highlighted with a white dashed line. \textcolor{black}{The microwave magnetic field $B_1(\omega_p)$ generated by the pump line is shown at some offset angle from $z$ because of flux-focussing effects.} In each figure the $B_0$ field orientation is displayed in the upper right corner.}
	\label{fig:chip}
\end{figure}

Central to the spectrometer is a co-planar waveguide resonator, fabricated from a film of superconducting \ch{NbN} sputtered onto a sapphire substrate and patterned using electron beam lithography into the `fractal' design shown in \cref{fig:chip} (a) \cite{DeGraaf2014}. The thicknesses of the \ch{NbN} and sapphire are \SI{140}{\nano\metre} and \SI{330}{\micro\metre} respectively.  The superconductor is $\SI{2}{\micro\metre}$ wide and separated by $\SI{2}{\micro\metre}$ at the resonator magnetic anti-node meaning that the microwave field $B_1$ is confined to a micron-scale mode volume as shown in the \textsc{Comsol} simulation in \cref{fig:chip} (d). The resonator is designed to be resilient to 
magnetic fields so that its high $Q$ is maintained up to $B_0\sim\SI{400}{\milli\tesla}$. 
It has the added benefit of a low flux focussing factor, meaning that 
orientation of $B_0$ must be broadly within the plane of the superconducting film of the resonator but time-consuming precise alignment is unnecessary. The resonator is capacitively coupled to a microwave feedline \cref{fig:chip} (b); the transmission across this line $S_{21}$ is measured to read out the amplitude $|S_{21}|$ and phase $\phi(S_{21})$. The resonator thereby acts as a very narrow notch filter with poor coupling to the spins when off-resonant.
 
 The spin sample is aligned on the resonator surface as shown in the image in  \cref{fig:chip} (a) at the magnetic anti-node of the $B_1$ field. Precise alignment is achieved with the 
 quartz tip of a micro-manipulator and the crystal is secured in place with a small amount of ESR-silent, cryogenic vacuum grease (Apiezon N-type). 
The sample is mounted on a printed circuit board (PCB) where a second microwave feedline (co-planar waveguide type of central strip width \SI{290}{\micro\metre}) lies underneath the resonator chip (therefore \SI{330}{\micro\metre} below the superconducting film) as demonstrated schematically in \cref{fig:chip}(b). This `pump' line serves as a second means of inputting microwave energy to the spin sample on the resonator. 

The PCB is mounted in a thermally anchored \ch{Cu} sample holder which is cooled to $\SI{65}{\milli\kelvin}$, the base temperature of our dilution refrigerator. Frequency-swept $S_{21}(\omega)$ transmission measurements are obtained using a vector network analyser (VNA) and a standard fit to the resonator lineshape \cite{Khalil2012} is used to extract the resonance frequency $\omega_0/2\pi = \SI{5.51}{\giga\hertz}$ and quality factor $Q=\num{3.3e4}$ at $\SI{198}{\milli\tesla}$ \cite{phaseExplanation}. 


The spin ensemble used in this work is a single microcrystal of the organic radical cation shown in \cref{fig:chip}(c) with a \ch{PF6^-} anion, which can be formed in good yield from electrocrystallisation in \ch{PhCl} with a \ch{Bu4NPF6} electrolyte \cite{Christensen2014}. The spin Hamiltonian for this system is nearly isotropic \textcolor{black}{$g\sim 2$}. Each molecule contains a single unpaired electron $S=\nicefrac{1}{2}$ corresponding to a single spin-allowed transition between the $m_s = \pm \nicefrac{1}{2}$ states. 
 
For ESR purposes, this spin sample is unusual in having a narrow linewidth (\SI{2.4}{\mega\hertz} full width at half maximum (FWHM) measured at $T=\SI{300}{\milli\kelvin}$) in spite of its high spin density  
\cite{Christensen2014}. This is attributed to the delocalisation of the unpaired electron across the extended $\pi$-system of the radical which means that electron-electron interactions are minimised, together with low inhomogeneous broadening because of the crystalline spin environment. 

\subsection{Microwave Circuitry}
\begin{figure}
\centering
	\includegraphics[width=\linewidth]{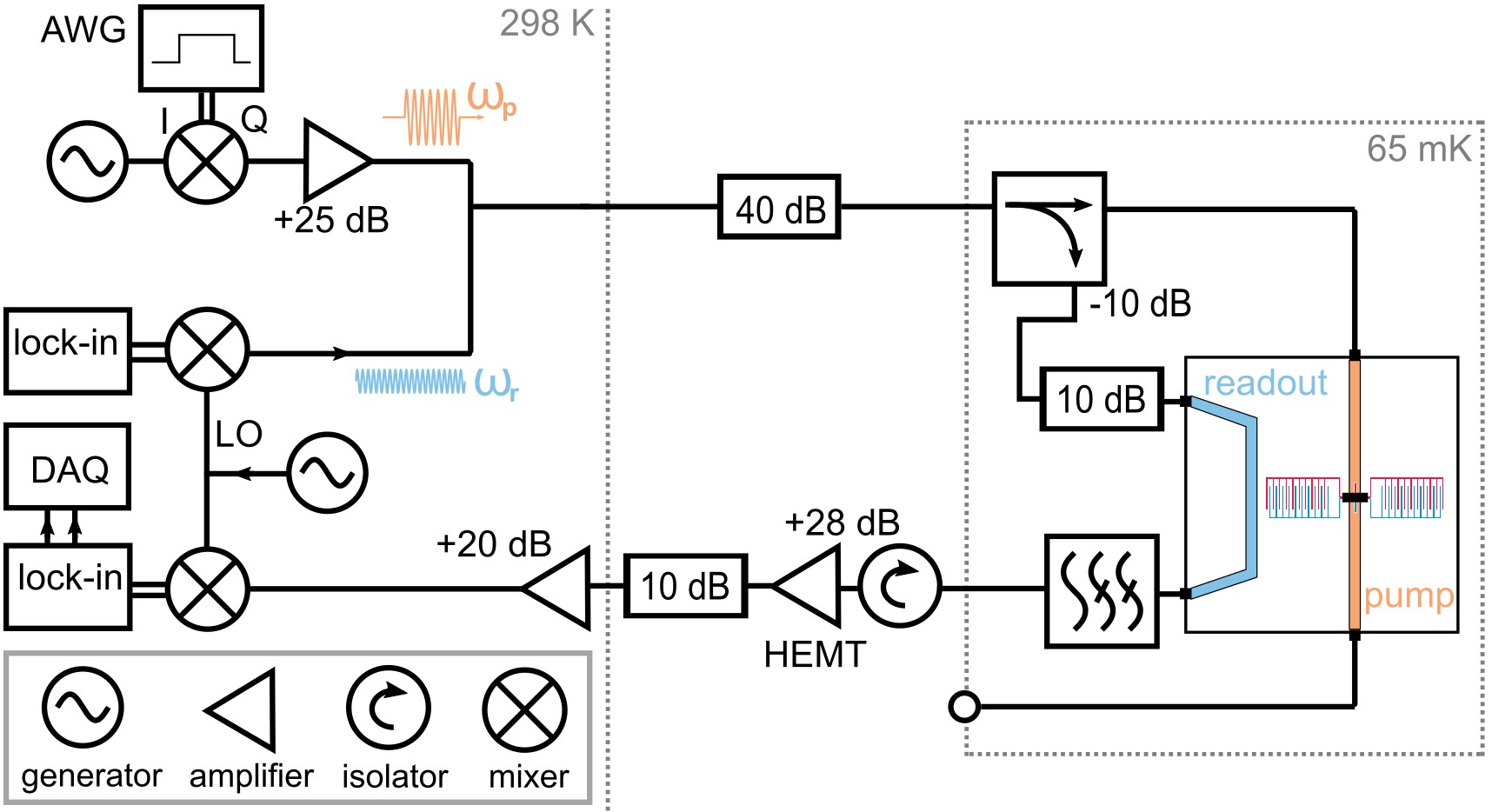}	
	\caption{Schematic of equipment: Heterodyne measurements of the resonator microwave transmission $S_{21}$ are implemented using a continuous wave microwave generator input to a single sideband mixer biased using a two-channel lock-in amplifier (each channel depicted as a separate box) to generate a single up-converted 
	sideband at $\omega_r$. The signal passes along the heavily attenuated fridge input line so that it is thermalised to the \SI{65}{\milli\kelvin} base temperature stage. It passes along a transmission line on the chip via a \SI{-10}{dB} coupler and further \SI{10}{dB} of attenuation. 
	The transmission from the chip is filtered and amplified at \SI{4}{\kelvin} and at room temperature before being demodulated by an IQ mixer and the second lock-in amplifier channel, from which data acquisition (DAQ) of the $I(t)$ and $Q(t)$ signals occurs.
	A second microwave generator produces a higher power tone of frequency $\omega_{p}$ shaped into pulses via an IQ mixer driven with the analog output of an AWG. The higher power used for $\omega_{p}$ means that it generates a significant radial $B_1$ field at the crystal when it passes along the pump line on the PCB. }
	\label{fig:mixers}
\end{figure}

An overview of the spectrometer setup is shown in \cref{fig:mixers} in which the room temperature microwave electronics are connected by a heavily attenuated input line to the readout and pump lines on the resonator chip. A heterodyne circuit is used to detect the amplitude and phase response of resonator at the readout frequency $\omega_r$. Separately, high power pulses are generated at frequency $\omega_p$ to probe the spin ensemble.

A microwave generator outputs a local oscillator (LO) signal which is up-converted 
 by \SI{40}{\mega\hertz} to $\omega_r$ at a single sideband mixer using a two-channel lock-in amplifier (Zurich Instruments model HF2LI). 
The phase of the output from the lock-in is tuned to select
exclusively the lower sideband from the mixer (so that any leakage in the upper sideband is remote from the spins which are at lower frequencies). 
This input tone receives \SI{60}{dB} of attenuation before it reaches the sample holder. 
The transmission signal from the microwave readout line is filtered and then amplified at \SI{4}{\kelvin} by a high-electron-mobility-transistor (HEMT) low noise amplifier. An isolator protects the sample from amplifier thermal radiation and signal reflections. 
Further amplification at room temperature precedes demodulation at the original LO frequency to produce a down-converted signal at \SI{40}{\mega\hertz} which is digitised, filtered and amplified at the lock-in amplifier. The resulting $I(t)$ and $Q(t)$ quadratures are acquired from an oscilloscope (Keysight model MSO-3014A), and the resonator transmission is then encapsulated in $S_{21}(t) = I(t) + i Q(t)$.

A second microwave generator produces a continuous signal at $\omega_{p}$. An AWG (Tektronix  5014C) with a nanosecond rise time is used to drive an IQ mixer and thereby modulate this generator output into arbitrary shaped pulses. These pulses are used to manipulate the spins temporally and independently of the resonator detection circuit. 
The spin transition frequency is designated $\omega_s$; pumping the spins resonantly corresponds to $\omega_p=\omega_s$. The pulses are amplified at room temperature, resulting in a power difference at the fridge input between the $\omega_{r}$ and $\omega_{p}$ signals in excess of \SI{55}{dB}.

The directional coupler immediately prior to the sample holder diverts $\nicefrac{9}{10}$ of the signal along the pump line. The $B_1$ field resulting from the pump line \textcolor{black}{has a radial dependence and moreover the flux-focussing effect of the patterned superconducting film distorts the field such that there are components from the pump line which are perpendicular to the $B_0$  as highlighted in \cref{fig:chip} (d). These components are necessary to drive the spins at the Zeeman transition. 
The magnitude of the field from the pump line is much smaller than that generated by the resonator if the two are compared at resonance $\omega_0$. However the pump line can be used to probe the spins when the resonator is significantly detuned and in this situation the resonator field resulting from probing the resonator has a negligible impact upon the spins}. 


\subsection{Dispersive Readout}
In advanced superconducting qubit architectures, dispersive readout is the standard technique for detecting the qubit state via a transmission measurement of the resonator \cite{Krantz2019,Arute2019,Wallraff2005}. When the two systems are detuned from each other by frequency $\Delta/2\pi$ which is greater than both the qubit-cavity coupling $g_k$ and the cavity decay rate $\kappa$, $\Delta\gg\{g_k,\kappa\}$, they reach the so-called `dispersive regime' in which the qubit causes a shift in resonator frequency which is dependent on the qubit state \emph{without} a direct exchange of energy between the two systems. This is a QND measurement because the outcome of the measurement is nominally unaltered by the measurement itself. A dispersive frequency shift of $\chi = g_{k}^2/\Delta$ will be observed so long as the number of photons in the resonator does not exceed the limit set by the critical photon number 
\textcolor{black}{which scales with the square of the detuning and with the inverse square of the cavity-qubit coupling} \cite{Krantz2019}. 
\textcolor{black}{This threshold is not the same} for an inhomogeneous ensemble of spins, where there is energy diffusion which requires more energy to be pumped into the system before saturation is observed. However, the same principle of dispersive readout in our system means that an $S_{21}$ measurement of the resonator can be used to infer the state and dynamics of the spin ensemble without interfering with the spins themselves. Even so, if the energy used to probe the resonator is too high, the resulting back-action interference with the spins reduces their spin-lattice coherence time \cite{Amsuss2011}.

Whereas the Jaynes-Cummings Hamiltonian models the single qubit (or spin)-resonator mode interaction described above \cite{Walls2008}, for our system this must be extended to encompass an ensemble of $N$ spins coupled to the resonator mode, which is treated by the Tavis-Cummings Hamiltonian \cite{Tavis1968,Garraway2011}. Further terms are added to model a pulse of frequency $\omega_p$ and amplitude $\eta(t)$ driving the spins. Each spin is described by a frequency $\omega_k$ and the Pauli operators $\{\sigma_k^{\pm},\sigma_k^z\}$; and $a(a^\dagger)$ is the standard annihilation (creation) operator for the cavity mode. It is now the ensemble coupling $g_{\textrm{ens}}=\sqrt{\sum_k^N g_k^2}$, and the spins' mean transition frequency $\omega_s$ which must be considered. The dispersive approximation \cite{Blais2004,Blais2020} applied to the Tavis-Cummings model \cite{Zueco2009} yields a Hamiltonian for the system which is valid if $\Delta = |\omega_0-\omega_s| \gg \{g_{\textrm{ens}},\kappa\}$.
\begin{equation}
\begin{split}
H  = \hbar a^\dagger a \Bigg(\omega_r + \sum_{k}^N \frac{g_k^2\sigma_k^z}{\Delta} \Bigg) 
+\frac{\hbar}{2} \sum_k^N \Bigg(\omega_k + \frac{g^2_k}{\Delta} \Bigg) \sigma_k^z \\
  -i \Bigg( 
\eta (t) a^\dagger e^{-i\omega_p t}
-\eta (t)^* a e^{i\omega_p t}
\Bigg)
\end{split}
\label{eq:disp_h}
\end{equation}
From the first term it can be seen
that the resonator frequency acquires a shift which is dependent upon the average spin polarisation $\langle\sum_k \sigma_k^z\rangle$. 
Evaluating the average spin polarisation in response to the pulsed drive requires numerical evaluation of the cavity-modified Bloch equations \cite{Bianchetti2009}. 

In measurements below, we detect changes to the phase of the resonator transmission in response to the changing spin polarisation in two contrasting situations: firstly a `linear' regime, where the phase of the resonator transmission is proportional to its frequency shift and thereby the spin polarisation; and secondly a `non-linear' regime where a greater dispersive shift means that the measurement of the phase cannot be used to directly infer the resonator frequency shift.
\section{Experimental results}
\label{exp}

\subsection{Continuous Wave (CW) ESR}
\begin{figure}[t]
	\includegraphics[width=0.99\linewidth]{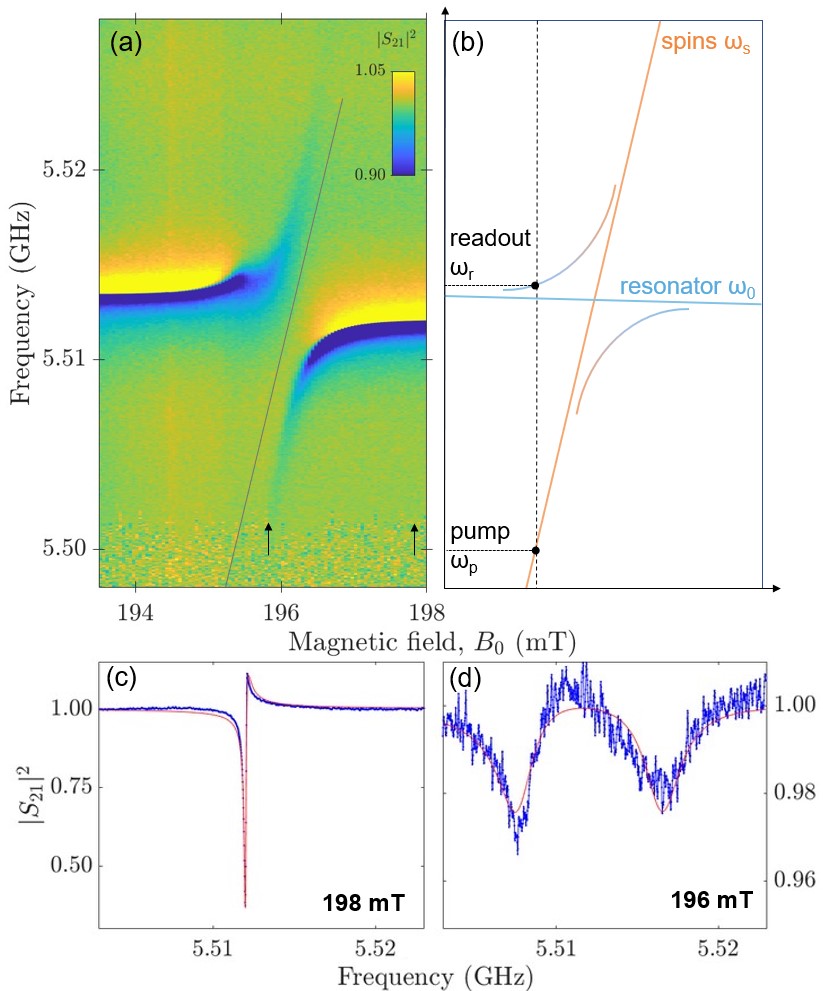}
	\caption{(a) The CW transmission as a function of magnetic field $|S_{21}(\omega,B_0)|^2$ which displays an avoided crossing where the spins are co-resonant with the resonator and (b) schematically to highlight the pump and readout tones used to interact with the spins and resonator in subsequent pulsed measurements. Line-cuts from (a) show the transmission of the uncoupled resonator at \SI{198}{\milli\tesla} (c) with the fit (red) to extract the total cavity decay rate $\kappa/2\pi = \SI[separate-uncertainty]{170(4)}{\kilo\hertz}$; and (d) the double-peaked hybrid spectrum at \SI{196}{\milli\tesla} and the fit (red) to \cref{eq:schuster} to extract $g_{\textrm{ens}}/2\pi = \SI[separate-uncertainty]{4.52(5)}{\mega\hertz}$. 
	}
	\label{fig:colourmap}
\end{figure}
To characterise the coupling between the resonator and the microcrystal, the VNA is connected across the fridge input line and the room temperature amplifier output in \cref{fig:mixers}. The microwave transmission through the readout line is measured as the magnetic field $B_0$ is swept to tune the Zeeman transition of the microcrystal spins $\omega_s$ through the resonator frequency $\omega_0$.

We use a probing microwave power of \SI{-90}{dBm} which corresponds to \num{e5} microwave photons in the resonator. The resulting plot in \cref{fig:colourmap}(a) displays the avoided crossing which is a signature of coherent coupling between the resonator and the spin ensemble. The spectrum can be calculated in full using the input-output formalism \cite{Clerk2010} 
\begin{equation}
	|S_{21}|^2 = \Biggl| 1+\frac{\kappa_c}{i(\omega-\omega_0) - (\kappa_c+\kappa_i) + \Big(\frac{|g_{\textrm{ens}}|^2}{i(\omega - \omega_s) - \gamma^*_2/2
	}\Big)}\Biggr|^2
	\label{eq:schuster}
	\end{equation}
	for a spin ensemble of linewidth $\gamma^*_2$ where the individual coupling to the spins $g_{k} \propto B_1$ \cite{phaseExplanation2}. 
	Here $\kappa_i$ and $\kappa_c$ are the internal and external cavity decay rates respectively. The 
	$|S_{21}|^2$ of the resonator transmission when detuned from the spins is shown in \cref{fig:colourmap}(c) and from the fit \cite{Khalil2012} we extract the cavity dissipation rate $\kappa = \kappa_i + \kappa_c = \omega_0/Q = 2\pi\times\SI[separate-uncertainty]{170(4)}{\kilo\hertz}$. 
	This value is then fixed to fit \cref{eq:schuster} to the avoided crossing data at $B_0=\SI{196}{\milli\tesla}$ when the resonator and spins are resonant in \cref{fig:colourmap}(d) and so obtain $g_{\textrm{ens}}/2\pi = \SI[separate-uncertainty]{4.52(5)}{\mega\hertz}$. A simulation of the expected resonator-spin coupling yields $g_{\textrm{ens}}/2\pi = \SI{8}{\mega\hertz}$ (average coupling strength per spin $g_0 = \SI{5}{\hertz}$) assuming that the crystal is a cuboid and sits almost flush with the resonator surface as depicted in \cref{fig:chip}(d). In fact the crystal is not perfectly cuboid-shaped, and is likely separated from the resonator surface by a small amount of vacuum grease both of which result in a small overestimation of the measured value. 
	
	The linewidth is more reliably extracted from the data in \cref{fig:linewidth}(b) as $\gamma_2^*/2 \pi = \SI[separate-uncertainty]{6.2(1)}{\mega\hertz}$: the coupling strength and linewidth are comparable and both exceed the resonator dissipation rate. From this we can evaluate the cooperativity $C = g_{\textrm{ens}}^2/\kappa\gamma_2^{*} \sim 19>1$ i.e. we have high-cooperativity but we do not obtain strong coupling, which requires $g_{\textrm{ens}}>\{\kappa,\gamma_2^*\}$.
		
\subsection{Pulsed ESR: Dispersive measurements in the linear regime}

The VNA measurement characterises the resonator-ensemble interaction as $B_0$ is incremented.
Here we set $B_0$ to \SI{193.5}{\milli\tesla} such that the resonator-spin detuning is $\Delta/2\pi = \SI{76}{\mega\hertz}\gg \{g_{\textrm{ens}},\kappa\}$. At this point, the resonator frequency shifts a small amount depending on the average spin polarisation. Because the phase of the resonator transmission is proportional to its frequency shift, we refer to this as the `linear' regime. In a traditional setup, the detuning would remove the means of exciting the spins whereas here the pump line allows us that freedom.
\begin{figure}[t]
	\includegraphics[width=\linewidth]{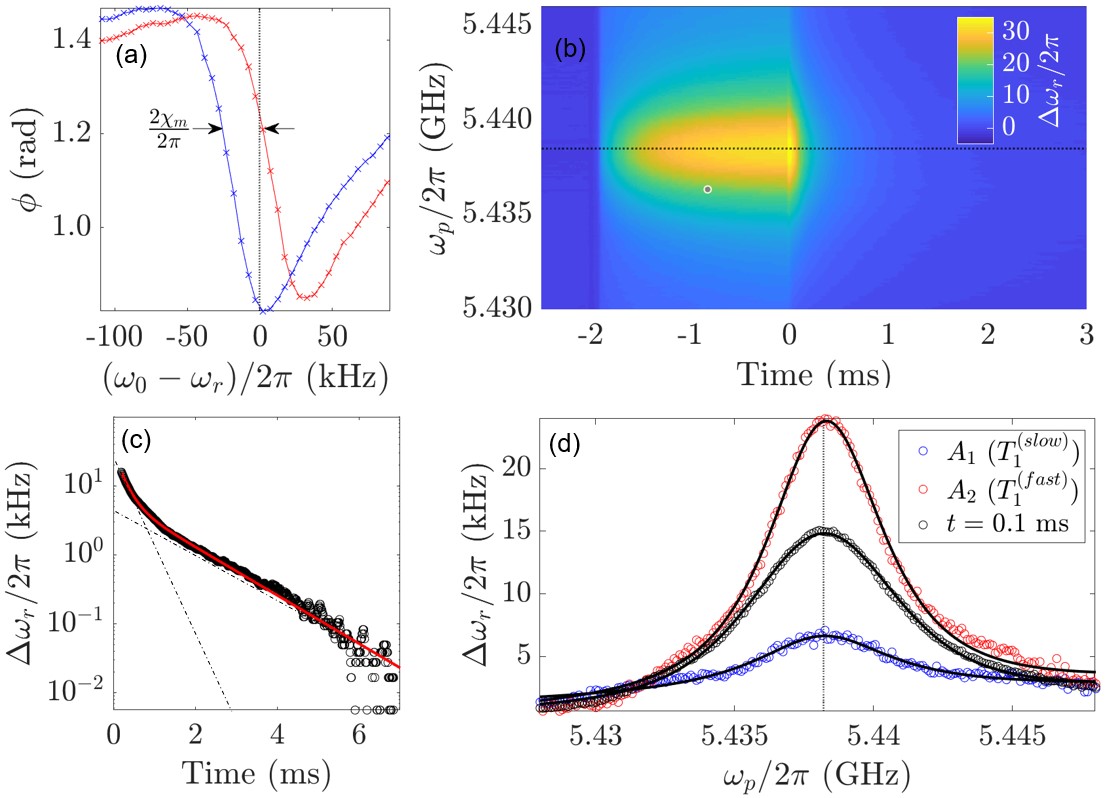}
	\caption{At $B_0=\SI{193.5}{\milli\tesla}$ (a) the resonator $\phi(S_{21})$ undergoes a dispersive shift in response to a high power CW pump tone applied at $\omega_p=\omega_s$ (blue trace) relative to the $\phi(S_{21})$ under a non-resonant pump tone at $\omega_p=\omega_s-2\pi\times\SI{10}{\mega\hertz}$ (red trace) resulting in a measured shift of $\chi_{\textrm{m}}/2\pi = \SI{15}{\kilo\hertz}$ where a sensible selection for the probe frequency $\omega_r$  is marked with a dashed vertical line 
	and (b) the frequency shift of the resonator $\Delta\omega_r$ in response to a \SI{2}{\milli\second} pulse applied at frequencies across the range $\omega_s-\SI{10}{\mega\hertz}<\omega_p<\omega_s+\SI{10}{\mega\hertz}$. A bi-exponential fit to the saturation recovery at $\omega_p=\omega_s$ after the pulse has terminated (c) extracts the time constants $T_1^{(\textrm{fast})}=\SI[separate-uncertainty]{0.23(4)}{\milli\second}$ and $T_1^{(\textrm{slow})}=\SI[separate-uncertainty]{1.24(5)}{\milli\second}$. The variation in the amplitude $A_1$ and $A_2$ of the two exponential terms as a function of $\omega_p$ reveals (d) q-Gaussian and Lorentzian terms from which the linewidths $\gamma_q^{(\textrm{fast})}/2\pi = \SI{4.9}{\mega\hertz}$ $(q=1.7)$ and $\gamma_q^{(\textrm{slow})}/2\pi = \SI{6.2}{\mega\hertz}$ $(q=2.1)$ are extracted; a cross section from (b) at \SI{0.1}{\milli\second} has linewidth $\gamma_2^*/2\pi = \SI{6.2}{\mega\hertz}$ $(q=1.4)$.}
	\label{fig:linewidth}
\end{figure}

 The microwave circuitry described in \cref{fig:mixers} supplies two microwave tones. A CW low power tone is applied to the resonator at frequency $\omega_{r}$ 
 whilst simultaneously a CW high power pump tone is applied either at $\omega_{s}$ or $\omega_{s}-2\pi\times\SI{10}{\mega\hertz}$. 
 When the pump tone is resonant with the spins, it diminishes their polarisation and causes a dispersive shift in the resonator frequency. As shown in \cref{fig:linewidth}(a), a frequency shift of $\chi_m=2\pi\times\SI{15}{\kilo\hertz}$ is observed in the phase response of the resonator when the pump is resonant with the spins (blue trace) as compared with the non-resonant pump (red trace). 
 This is less than the maximum $\chi = g_{\textrm{ens}}^2/\Delta = 2\pi\times\SI{270}{\kilo\hertz}$ \cite{Stammeier2017}, which suggests that the pump tone is not able to fully saturate the spins and therefore they do not exert the maximum frequency shift upon the resonator.
The measurement in \cref{fig:linewidth}(a) is a calibration step, which allows us to infer the frequency shift of the resonator from the phase of the $S_{21}$. As with qubit measurements, the readout frequency $\omega_r$ (marked by vertical dashed line) is chosen at the point where the dependence of the $\phi(S_{21})$ upon the spins' state is at a maximum \cite{Blais2004}. It is important that we know the gradient of the phase lineshape, and that there is a linear relation between the resonator phase behaviour and its frequency shift $\Delta \omega_r$ at the chosen readout frequency.
 
 Using this readout frequency, a pulsed measurement is obtained whereby a high-amplitude \SI{2}{\milli\second} pulse drives the resonant spin transition and a weak readout tone then detects the return to equilibrium polarisation. The pump pulse is swept across the spin linewidth in the range $\omega_p = \omega_s\pm 2\pi\times\SI{10}{\mega\hertz}$. The resulting spectral density plot in \cref{fig:linewidth}(b) shows the temporal response of $\Delta\omega_r$ to the saturating pulse as a function of $\omega_p$. A cross-section in \cref{fig:linewidth}(d) at \SI{0.1}{\milli\second} after the pulse terminates reveals the spin linewidth to which we find good agreement with a pseudo-Gaussian distribution fit \cite{Krimer2014} of FWHM $2\pi\times\SI[separate-uncertainty]{6.2(1)}{\mega\hertz}$ $(q = 1.40)$. The dimensionless parameter $q$ describes the lineshape: $q\rightarrow 1$ ($q = 2$) obtains a pure Gaussian (Lorentzian) distribution. 


The decay of the spins is examined by fitting $\Delta\omega_r(t)$ after the saturating pulse terminates and the spin polarisation is left to return to equilibrium. The decay at $\omega_p=\omega_s$ is shown in \cref{fig:linewidth}(c); it is bi-exponential $A_1 \exp(-t/T_1^{\textrm{(fast)}})+A_2 \exp(-t/T_1^{\textrm{(slow)}})$ with time constants \SI[separate-uncertainty]{0.23(4)}{\milli\second} and \SI[separate-uncertainty]{1.24(5)}{\milli\second}. 

By fitting the polarisation decay to a bi-exponential law at different $\omega_p$ in \cref{fig:linewidth}(d) we observe that the amplitude of the decay terms $A_1(\omega_p)$ and $A_2(\omega_p)$ follow two different distributions. For the fast $T_1$ the linewidth has FWHM \SI[separate-uncertainty]{4.9(2)}{\mega\hertz} $(q=1.7)$, whereas for the slow $T_1$ there is FWHM of \SI[separate-uncertainty]{6.2(5)}{\mega\hertz} with $q=2.0$ and thus the distribution is almost Lorentzian. The different lineshape properties could suggest that the two relaxation times originate from different paramagnetic species \cite{Ranjan2013}. However the likely spin contaminant from within the material of the resonator has an entirely different linewidth \cite{DeGraaf2017}, suggesting that the two observed relaxation times may instead arise from differences within the crystal itself. 

\subsection{Pulsed ESR: Dispersive measurements in the non-linear regime} 
\begin{figure}[t]
\centering
	\includegraphics[width=\linewidth]{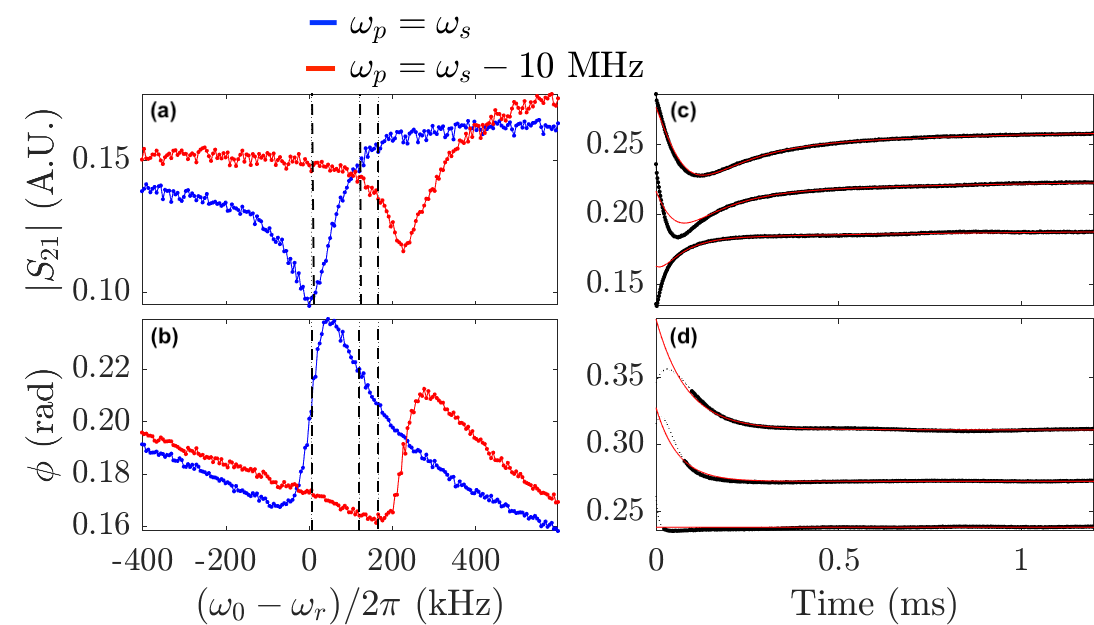}
	\caption{At \SI{195}{\milli\tesla} the resonator mode dispersively shifts by \SI{230}{\kilo\hertz} as shown in amplitude (a) and phase (b) data when the pump tone is resonant with the spins $\omega_p=\omega_s=2\pi \times \SI{5.48}{\giga\hertz}$ (blue) compared to when the pump tone is non-resonant (red). 
	When the readout frequency $\omega_r$ is fixed to the three values marked by vertical dashed lines in (a) and (b), and the $\SI{2}{\milli\second}$ pump pulse is set to $\omega_p=\omega_s$, the spin dynamics can be inferred via the amplitude (c) and phase (d) of the resonator $S_{21}$ as the system equilibrates (traces offset for clarity). The amplitude fit at all the measured $\omega_r$ extracts $T_1^{(\textrm{slow})}=\SI[separate-uncertainty]{1.0(2)}{\milli\second}$ and $T_1^{(\textrm{fast})}=\SI[separate-uncertainty]{0.11(5)}{\milli\second}$. The latter portion of the phase decay tracks a linear region of the resonator lineshape in phase (b) and so a straightforward bi-exponential fit (red) is enforced in (d) using the same time constants.}
	\label{fig:pulse-shift}
\end{figure}

Tuning the field to $B_0=\SI{195}{\milli\tesla}$ diminishes the resonator-spin detuning to $\Delta/2\pi=\SI{32}{\mega\hertz}$. A higher power $\omega_p$ tone is also used; the combined effect is that the maximum possible shift increases $\chi/2\pi = \SI{630}{\kilo\hertz}$. Again, due to insufficient power to saturate the spins the observed shift is $\chi_m/2\pi=\SI{120}{\kilo\hertz}$. As can be seen in the calibration measurement \cref{fig:pulse-shift}(b) this means that there is no longer a region where a readout frequency $\omega_r$ can be selected such that the change in phase of resonator transmission is linearly proportional to its frequency shift, which we refer to as `non-linear' in terms of the resonator readout. Being able to use the resonator as a probe to the spins in spite of this increases the range of dispersive measurements which can be taken. 

Pulsed measurements in the non-linear regime were acquired as before by applying a high-power \SI{2}{\milli\second} saturating pulse at  $\omega_p=\omega_s$ and detecting the decay in the resonator $\phi(S_{21})$ \cref{fig:pulse-shift}(d) and $|S_{21}|$ \cref{fig:pulse-shift}(c).
To fit the data, \cref{eq:schuster} is modified to incorporate the time-dependence of the ensemble polarisation after the saturating pulse has terminated into the coupling term, where $P$ is a fit parameter to account for the incomplete polarisation of the ensemble 
\begin{equation}
		G(t) = g_{\textrm{ens}}P
		\big(1-A_1 e^{-t/T_1^{\textrm{(fast)}}}-A_2 e^{-t/T_1^{\textrm{(slow)}}}\big).
	\label{eq:g_time}	
\end{equation}
Inserting this expression into \cref{eq:schuster} gives a good fit to $|S_{21}(t)|$,
examples of which are shown in \cref{fig:pulse-shift}(c) and we extract $T_1^{(\textrm{slow})}=\SI[separate-uncertainty]{1.0(2)}{\milli\second}$ and $T_1^{(\textrm{fast})}=\SI[separate-uncertainty]{0.11(5)}{\milli\second}$ which are in reasonable agreement with the values obtained from the linear regime data. Larger uncertainty is to be expected given the increased number of variables in the fit. 
Towards the end of the phase decay the resonator dispersive shift will be so sufficiently diminished as to `re-gain' the linear regime; therefore in \cref{fig:pulse-shift}(d) a bi-exponential fit is enforced to the tail of the decay using the $T_1$ values from \cref{fig:pulse-shift}(c). 

Even at zero detuning, the spin-lattice relaxation rate \SI{500}{\kilo\hertz} is far greater than the Purcell rate for cavity-enhanced spontaneous emission \SI{4}{\milli\hertz} i.e. the system is far from being limited by Purcell emission in this case.  
\section{Conclusion}
We have demonstrated a setup and methodology for dispersive pulsed ESR of a pico-litre organic crystal using a superconducting resonator with high $Q$ and small mode volume. The \SI{4.5}{\mega\hertz} resonator-spin coupling causes the resonator frequency $\omega_0$ to dispersively shift in response to the spin polarisation. A separate pump line enables spin excitation irrespective of the detuning from the resonator frequency. Pulsed ESR saturation recovery measurements to detect the $T_1$ time of the sample made use of the two-tone dispersive readout. This allowed us to monitor the spin dynamics when the resonator-spin detuning is large (whereas no such detuning is available in conventional ESR), and there is minimal coupling between the spins and the $B_1$ field of the resonator itself. We show that saturation recovery measurements are consistent whether the resonator frequency is shifted within or beyond the linear regime, increasing the range of detuning which can be investigated with this method. 

The establishment of this measurement protocol brings an additional degree of freedom to ESR measurements: the resonator is a probe of the spin dynamics even when the two are substantially detuned from each other. Not only this but the minimisation of back-action upon the spins, which is inherent to a dispersive measurement, creates the possibility 
\textcolor{black}{of using this technique to readout the quantum coherent states of the spin ensemble when executing quantum information protocols}.

\section*{Acknowledgements}
The authors thank Mogens Br\o{}ndsted Nielsen and Christian R. Parker at the University of Copenhagen for providing a sample of the microcrystal used in this work, and Alexander Tzalenchuk for helpful suggestions and comments.
This project has received funding from the UK department for Business, Energy and Industrial Strategy through the UK national quantum technologies program. A.K. and M.O. gratefully acknowledge financial support from the EPSRC Training and Skills Hub in Quantum Systems Engineering EP/P510257/1 and the EPSRC Manufacturing Research Fellowship EP/K037390/1 respectively. S.K. and A.D. acknowledge support from the Swedish Research Council (VR) (grant agreements 2016-04828 and 2019-05480), EU H2020 European Microkelvin Platform (grant agreement 824109). The authors declare no competing financial interests.
\section*{References}



\bibliographystyle{elsarticle-num} 
\bibliography{assorted_experimental-danish-experimental-edit}

\begin{thebibliography}{10}
\bibitem{phaseExplanation}
{A phase rotation factor $e^{i\phi}$ is introduced to account for slight
  impedance mismatch along the feedline ($\phi = \SI{-0.6}{\radian}$ gives the
  best fit to the data).}
\bibitem{phaseExplanation2}
{As with the non-resonant fit to the resonator lineshape, a phase rotation
  factor $e^{i\phi}$ is introduced to account for slight impedance mismatch
  along the feedline ($\phi = \SI{-0.6}{\radian}$ gives the best fit to the
  data).}
\end{thebibliography}

\end{document}